\catcode`\@=11

\expandafter\ifx\csname draftcontrol\endcsname\relax\global\def\draftcontrol{0}
\fi

{\count255=\time\divide\count255 by 60
\xdef\hourmin{\number\count255}
\multiply\count255 by-60\advance\count255 by\time
\xdef\hourmin{\hourmin:\ifnum\count255<10 0\fi\the\count255}}
\def\draftdate{\number\month/\number\day/\number\year\ \ \ \hourmin }

\newcommand\makepapertitle{\par
  \begingroup
    \renewcommand\thefootnote{\@fnsymbol\c@footnote}%
    \def\@makefnmark{\rlap{\@textsuperscript{\normalfont\@thefnmark}}}%
    \long\def\@makefntext##1{\parindent 1em\noindent
            \hb@xt@1.8em{%
                \hss\@textsuperscript{\normalfont\@thefnmark}}##1}%
     \newpage
     \global\@topnum\z@   
     \@makepapertitle
     \thispagestyle{empty}\@thanks
  \endgroup
  \setcounter{footnote}{0}%
  \global\let\thanks\relax
  \global\let\makepapertitle\relax
  \global\let\@makepapertitle\relax
  \global\let\@thanks\@empty
  \global\let\@author\@empty
  \global\let\@date\@empty
  \global\let\@title\@empty
  \global\let\title\relax
  \global\let\author\relax
  \global\let\date\relax
  \global\let\and\relax
  \def\version{\let\version\@version\@gobble}
}
\def\@makepapertitle{%
  \newpage
   \ifnum\draftcontrol=1 {}
   \version\versionno
   \vskip 3em%
   \else
   \hfill\hbox to 3cm {\parbox{4cm}{\@pubnum}\hss}%
   \vskip 3em%
   \fi
   \begin{center}%
   \let \footnote \thanks
     {\LARGE \@title \par}%
     \vskip 1.5em%
     {\normalsize
       \lineskip .5em%
       \begin{tabular}[t]{c}%
         \@author
       \end{tabular}\par}%
     \vskip 1em%
     {\@bstract}%
     \end{center}%
     \vskip .5em
     \@date%
   \par
}

\gdef\@pubnum{}
\def\pubnum#1{%
  \gdef\@pubnum{#1}}

\gdef\@bstract{}
\def\Abstract#1{%
  \gdef\@bstract{%
   \parbox{\textwidth-0pc}{%
   \centerline{\bf Abstract}\penalty1000
   \noindent
   \renewcommand\baselinestretch{1.0}
   {#1}}}
}

\def\ps@paper{\let\@mkboth\@gobbletwo%
     \ifnum\draftcontrol=1
        \def\@oddfoot{\hbox to \textwidth{\tiny \versionno \hfil\tiny\draftdate}%
        \hskip -\textwidth \hbox to \textwidth{\hfil\rm\thepage\hfil}}%
     \else\def\@oddfoot{\hbox to \textwidth{\hfil\rm\thepage\hfil}}
     \fi
     \let\@evenfoot\@oddfoot
}

\def\body{\clearpage
          \pagestyle{paper}
        }
\newenvironment{acknowledgments}{%
\vskip 3.25ex
\noindent {\bf Acknowledgments}
}

\def\@version#1{\ifnum\draftcontrol=1
\typeout{}\typeout{#1}\typeout{}
\vskip3mm\centerline{\hbox{\fbox{\normalsize{\tt DRAFT -- #1 -- }
                   {\draftdate}}}}\vskip3mm
\fi}
\let\version\@version
\long\def\eqlabel#1{\ifnum\draftcontrol=1
                    \tag@false  
                    \tag*{(\theequation) \hbox to -0.2cm{\hspace{0cm}\small{#1}\hss}}
                    \refstepcounter{equation} 
                    \edef\@currentlabel{\theequation}
                    \ltx@label{#1}          
                    \else
                    \label{#1}
                    \fi
                    }
\let\st@bibitem\@bibitem
\let\st@lbibitem\@lbibitem
\ifnum\draftcontrol=1
  \def\@bibitem#1{%
    \st@bibitem{#1}\a@@label{#1}\ignorespaces}
  \def\@lbibitem[#1]#2{%
    \st@lbibitem[#1]{#2}\a@@label{#2}\ignorespaces}
  \def\a@@label#1{%
    \gdef\a@lab{\smash{\normalfont\small#1}}
    \ifvmode
      \if@inlabel
        \global\setbox\@labels\hbox{%
          \llap{\a@lab\let\a@lab\relax
                \kern\@totalleftmargin\kern\marginparsep}%
          \box\@labels}%
      \fi
    \fi}
\fi

\documentclass[12pt,letterpaper]{article}

\usepackage{amsmath,amssymb,array,calc,rotating,epsfig,psfrag}
\usepackage[nosort]{cite}

\ifnum\draftcontrol=1
\fi

\renewcommand\baselinestretch{1.25}
\renewcommand\section{\@startsection {section}{1}{\z@}%
                                   {-3.5ex \@plus -1ex \@minus -.2ex}%
                                   {2.3ex \@plus.2ex}%
                                   {\normalfont\large\bfseries}}
\renewcommand\subsection{\@startsection{subsection}{2}{\z@}%
                                     {-3.25ex\@plus -1ex \@minus -.2ex}%
                                     {1.5ex \@plus .2ex}%
                                     {\normalfont\normalsize\bfseries}}
\renewcommand\subsubsection{\@startsection{subsubsection}{3}{\z@}%
                                     {-3.25ex\@plus -1ex \@minus -.2ex}%
                                     {1.5ex \@plus .2ex}%
                                     {\normalfont\normalsize\it}}

\numberwithin{equation}{section}

\def\zet          {{\mathbb Z}}

\def\del          {\partial}

\def\ee		{{\rm e}}

\def\be		{\begin{equation}}
\def\ende		{\end{equation}}

\def\revise#1       {\marginpar{\rule{2mm}{1cm} #1}}

\def\ZZ{\zet}

\def\R{{\rm R}}

\def\sqr#1#2{{\vcenter{\vbox{\hrule height.#2pt  
 \hbox{\vrule width.#2pt height#1pt \kern#1pt
 \vrule width.#2pt}\hrule height.#2pt}}}}

\def\yboxit#1#2{\vbox{\hrule height #1 \hbox{\vrule width #1
\vbox{#2}\vrule width #1 }\hrule height #1 }}
\def\fillbox#1{\hbox to #1{\vbox to #1{\vfil}\hfil}}
\def\ybox{{\lower 1.3pt \yboxit{0.4pt}{\fillbox{8pt}}\hskip-0.2pt}}

\def\comments#1{}

\def\P{\BP}

\def\II{\relax{I\kern-.10em I}}

\def\IZ{\relax\ifmmode\mathchoice
{\hbox{\cmss Z\kern-.4em Z}}{\hbox{\cmss Z\kern-.4em Z}}
{\lower.9pt\hbox{\cmsss Z\kern-.4em Z}}
{\lower1.2pt\hbox{\cmsss Z\kern-.4em Z}}\else{\cmss Z\kern-.4em
Z}\fi}
\def\IB{\relax{\rm I\kern-.18em B}}
\def\IC{{\relax\hbox{$\inbar\kern-.3em{\rm C}$}}}
\def\ID{\relax{\rm I\kern-.18em D}}
\def\IE{\relax{\rm I\kern-.18em E}}
\def\IF{\relax{\rm I\kern-.18em F}}
\def\IG{\relax\hbox{$\inbar\kern-.3em{\rm G}$}}
\def\IGa{\relax\hbox{${\rm I}\kern-.18em\Gamma$}}
\def\IH{\relax{\rm I\kern-.18em H}}
\def\II{\relax{\rm I\kern-.18em I}}
\def\IK{\relax{\rm I\kern-.18em K}}
\def\IP{\relax{\rm I\kern-.18em P}}

\def\inbar{\,\vrule height1.5ex width.4pt depth0pt}

\font\cmss=cmss10 \font\cmsss=cmss10 at 7pt
\def\IR{\relax{\rm I\kern-.18em R}}

\def\BR{\IR}
\def\BP{\IP}
\def\BR{\IR}

\def\rp{\IR\IP}

\def\lp10{l_P^{10}}
\def\lp11{l_P^{11}}

\newcommand{\nc}{\newcommand}
\nc{\rnc}{\renewcommand}
\nc{\CY}{Calabi-Yau}
\nc{\CYM}{Calabi-Yau manifold}
\nc{\CYMs}{Calabi-Yau manifolds}
\nc{\DB}{D-Brane}
\nc{\DBs}{D-Branes}
\nc{\SUSY}{supersymmetry}
\nc{\Kah}{K\"ahler}
\nc{\cs}{complex structure}
\nc{\beq}{\begin{equation}}
\nc{\eeq}{\end{equation}}
\nc{\ntwo}{${\cal N}=2$}
\nc{\nOne}{${\cal N}=1$}
\nc{\hs}{\hspace{0.2in}}
\nc{\Z}{{\mathbb Z}}
\rnc{\P}{{\mathbb P}}
\nc{\WP}{\mathbb{WP}}
\nc{\slag}{special Lagrangian}
\nc{\cn}{\C^n}
\nc{\rn}{\R^n}

\nc{\SO}{\hbox{SO}}
\nc{\Sp}{\hbox{Sp}}
\nc{\SU}{\hbox{SU}}

\def\bea		{\begin{eqnarray}}
\def\eea		{\end{eqnarray}}
\nc{\e}{{\rm exp}}

\nc{\cosech}{{\rm cosech}}

\nc{\Li}{{\rm Li_{2}}}

\catcode`\@=12

\begin{document}

\title{Chern-Simons Matrix Models and Unoriented Strings.} 

\pubnum{%
USC-03-04 \\
hep-th/0305134}
\date{May 2003}

\author{Nick Halmagyi\footnote{halmagyi@physics.usc.edu} and Vadim Yasnov\footnote{yasnov@physics.usc.edu}\\[0.4cm]
\it Department of Physics and Astronomy\\
\it University of Southern California \\
\it Los Angeles, CA 90089, USA \\[0.2cm]
}

\Abstract{
For matrix models with measure on the Lie algebra of $SO/Sp$,
the sub-leading free energy is given by 
$F_{1}(S)=\pm\frac{1}{4}\frac{\del F_{0}(S)}{\del S}$. 
Motivated by the
fact that this relationship does not hold for Chern-Simons theory on $S^{3}$,
we calculate the sub-leading free energy in the matrix model for this theory, which is 
a Gaussian matrix model with Haar measure on the group $SO/Sp$.
We derive a quantum loop equation for this matrix model
and then find that $F_{1}$ is an integral of 
the leading order resolvent over the spectral curve. We explicitly calculate
this integral for quadratic potential and find agreement with previous studies of $SO/Sp$ Chern-Simons theory.
}

\enlargethispage{1.5cm}

\makepapertitle

\vfill \eject 

\tableofcontents

\body

\version\versionno

\section{Introduction}

Matrix models have been studied intensely since the classic paper \cite{Brezin:1977sv}, where it was realized that they enumerate planar diagrams.
Remarkably, it was recently discovered that there is a description of
Chern-Simons (CS) theory \cite{Marino:2002fk, Aganagic:2002wv}
and Holomorphic Chern-Simons (HCS) theory \cite{Dijkgraaf:2002fc, Dijkgraaf:2002vw, Dijkgraaf:2002dh} on certain manifolds as particular matrix models. Whilst interesting in its own right, this discovery
also has deep implications for supersymmetric gauge theory in four dimensions.

The connections between CS/HCS theory and field theory in four dimensions is the following (see \cite{Marino:2002wa} for a review). 
Type II string theory on a Calabi-Yau (CY) manifold
can be twisted in one of two ways to give a topological string, the so called A and B models \cite{Witten:1991zz}. Furthermore, it was shown in
\cite{Witten:1992fb} that the open string A-model on the CY manifold $T^{*}M$, where $M$ is a Lagrangian submanifold,
is equivalent to CS theory on $M$ with gauge group determined by the Chan-Paton factors. It was also shown that 
the open string B-model on a CY is equivalent to HCS theory on that CY. So calculating the partition function
of these CS/HCS theories amounts to solving the topological A/B string.

For certain topological correlators the A/B twist is trivial and thus these topological correlators give physical correlators, this in fact happens
for precisely the correlators that correspond to F-terms in the resulting
four dimensional gauge theory \cite{Witten:1991zz}. So the conclusion is that the superpotential can be
computed entirely from the topological string. This was made extremely precise in \cite{Bershadsky:1993cx, Dijkgraaf:2002dh},
where it was shown that the effective superpotential in four dimensions $W_{eff}$ is given by,

\be
W_{eff}=N\frac{\del F_{0}}{\del S}-\tau S,
\ende
where, N is the amount of RR flux, $F_{0}$ is the leading order contribution to the topological string free energy, $S$ is the gaugino condensate and $\tau $ is the gauge coupling. 
So showing that topological strings can be described by matrix models is very suggestive that the matrix model structure can be uncovered directly in field theory. 
For the case of $SU(N),\ {\cal N}=1$ gauge theory with adjoint matter,
this was found in \cite{Dijkgraaf:2002xd, Cachazo:2002ry}.

After the initial work of Dijkgraaf and Vafa, it was shown how to generalize
their work to the other classical gauge groups from several points of view. Matrix model generalizations were considered in 
\cite{Ashok:2002bi, Janik:2002nz}, the
perturbative supergraph techniques of \cite{Dijkgraaf:2002xd} were considered in \cite{Ita:2002kx} and generalized Konishi anomaly techniques of \cite{Cachazo:2002ry} were considered in 
\cite{Alday:2003gb, Kraus:2003jv}. All these works were studying ${\cal N}=1\  SO/Sp$ gauge theories in four dimensions with adjoint matter and single trace superpotential.

The present work is concerned with how the calculations of \cite{Ashok:2002bi} (which will be reviewed in section 2) can be performed in the CS model matrix model of \cite{Marino:2002fk, Aganagic:2002wv}. 
The main result of this paper is the calculation of $F_{1}$ in the CS theory on $S^{3}$ with gauge group $SO/Sp$. The partition function of this CS theory has in fact been calculated to all orders in \cite{Sinha:2000ap}, the present work explores by explicit calculation, the matrix model description of CS theory.
At first glance there appears to be an inherent contradiction between a naive extrapolation of the results of \cite{Ashok:2002bi} and the known partition function \cite{Sinha:2000ap}.
We will find in this paper that although the CS
case is more complicated than the Lie algebra matrix model, the matrix model realizations of CS
is not incorrect.

In \cite{Aganagic:2002wv}, it was shown that the CS model matrix model is of the same type as the B-model matrix model but with a rather complicated double trace potential. It must be because of the double trace that the B-model calculation does not translate to the CS model.

This paper is organized as follows. In section 2 we review the calculation of \cite{Ashok:2002bi} and present
a new way to obtain the same results. This new method will generalize to the CS matrix model. In section 3 we describe
the topological string and Calabi-Yau geometry which is inherently being studied. 
Section 4 will contain a discussion of the free energy of $SO/Sp$ CS theory, what the naive contradiction 
is and what we will calculate from the matrix model. Section 5 will contain a derivation
of the loop equation for matrix models with Haar measure, something which has not appeared in the literature. 
We will see explicitly why the method of \cite{Ashok:2002bi} breaks down.
In section 6 we will calculate the leading order free energy of the CS matrix model for groups $SO/Sp$, this is a straightforward generalization of \cite{Aganagic:2002wv} but is included for completeness. 
In section 7 we will calculate the subleading free energy,
this is the main result of the paper. In section 8 we will discuss the four dimensional gauge theory 
which this analysis actually corresponds to.

\section{Matrix Models for Classical Groups}
Matrix models for all the classical groups were first considered in \cite{Myers:pp}, where they wrote down the appropriate measures in eigenvalue form. More recently, the authors of \cite{Ashok:2002bi} studied the orientifolded CY geometry of \cite{Dijkgraaf:2002fc} and the resulting matrix model. As is well known, orientifolding internal geometries in string theory leads to the gauge groups $SO/Sp$. It was shown in \cite{Ashok:2002bi} that the relevant matrix model, when in eigenvalue form, has a measure on the Lie algebra of $SO/Sp$, as considered
previously in \cite{Myers:pp}. 

\subsection{A first look at $F_{1}$}

Techniques for calculating ${\cal O}(N^{-2})$ and higher corrections to the Hermitian matrix model free energy were considered previously in \cite{Ambjorn:1992gw, Akemann:1996zr}(they correspond to four dimensional gravitational F-terms). The same method was used in \cite{Ashok:2002bi} to derive the ${\cal O}(N^{-1})$ correction to the free energy ($F_{1}$) and will now be reviewed. First, lets set up some notation. The partition function of the matrix model is,

\be
{\cal Z}=\int d\Phi \ee ^{-\frac{1}{g_s}{\rm Tr}W(\Phi)},
\ende
where, $W=\sum_{j=1}^{\infty}\frac{g_{j}}{2j}\Phi^{2j}$, and $\Phi$ is in the adjoint of $SO/Sp$.
For a single cut model, the number of eigenvalues in that cut is $M$, and we define $S=g_{s}\frac{M}{2}$.
The resolvent is defined as, (g is for genus, c is for crosscap)

\be
\omega(x)=g_{s}\left\langle {\rm Tr}\frac{1}{x-\Phi} \right\rangle=\sum_{g,c}g_{s}^{2g+c}\omega_{2g+c}(x)
\ende
Now we need the relation,

\be
\omega(x)=\frac{d}{dV}(x)F+\frac{S}{x}
\ende
where $\frac{d}{dV}(x)=-\sum_{j=1}^{\infty}\frac{2j}{x^{2j+1}}\frac{\del}{\del g_{j}}$,
and the $g_{s}$ expansion of $F$, namely 
\be
F=\sum_{g,c}g_{s}^{2g+c}F_{2g+c},
\ende
to see that

\bea\label{om0om1}
\omega_{0}=\frac{d}{dV}(x)F_{0}+\frac{S}{x} \\
\omega_{1}=\frac{d}{dV}(x)F_{1}. \nonumber
\eea
Then we derive a loop equation for the full resolvent, the answer being

\be \label{le1}
2\oint_{C}\frac{dx^{'}}{2\pi i}\frac{W^{'}(x^{'})}{x-x^{'}}\omega(x^{'})
=\omega(x)^{2} - \frac{g_{s}}{x}\omega(x)+g_{s}^{2}\frac{d}{dV}(x)\omega(x),
\ende
and using the $g_{s}$ expansion of $\omega(x)$, we extract the zeroth and first order loop equations,

\bea 
2\oint_{C}\frac{dx'}{2\pi i}\frac{W'(x')}{x-x'}\omega_{0}(x')
=\omega_{0}(x)^{2}, \label{exp_loop} \\
2\oint_{C}\frac{dx'}{2\pi i}\frac{W'(x')}{x-x'}\omega_{1}(x')
=2\omega_{1}(x)\omega_{0}(x)-\frac{1}{x}\omega_{0}(x). \label{g_s_loop}
\eea
Now one observes that a solution for $\omega_{1}(x)$ is given by,

\be \label{om1}
\omega_{1}(x)=\frac{1}{2x}-\frac{1}{4}\frac{\del \omega_{0}}{\del S}
\ende
then using (\ref{om0om1})we see that this implies

\be \label{free1}
F_{1}=-\frac{1}{4}\frac{\del F_{0}}{\del S}.
\ende
And so we have the first correction to the free energy in terms of the planar free energy, valid for any single trace potential. 

\subsection{A second look at $F_{1}$}

The above method of obtaining $F_{1}$ does not generalize to the case 
of CS matrix models, as will become evident in section 5.
So we will now find $F_{1}$ 
for this model using a rather different method than the one above, one which will
generalize to the case of the CS matrix model.

We will assume there is just a single cut $(-a,a)$.
We then modify the potential as follows,

\be
\label{addW}
\widehat{W}(x) = W(x) - \frac{g_s}{2}{\rm ln}(x).
\ende
We will denote the free energy of this modified model
$\widehat{F}$. This modification effects the 
loop equation (\ref{le1}), by adding to the LHS, a term

\be
\frac{g_{s}}{2}\oint_{{\cal C}}\frac{dx'}{2\pi i} \frac{\omega(x')}{x'(x-x')},
\ende
where as usual, the contour does not encircle the point $x$.
Now by deforming the contour to infinity we pick up the point $x$ but the
integral around infinity vanishes and we see that 

\be
\oint_{{\cal C}}\frac{dx'}{2\pi i} \frac{\omega(x')}{x'(x-x')}= \frac{\omega(x)}{x}.
\ende

So modifying the potential in this way leaves the leading order loop equation (\ref{exp_loop}) unchanged,
but the ${\cal O}(g_{s})$ loop equation (\ref{g_s_loop}) becomes,

\be
\oint_{C}\frac{dx'}{2\pi i}\frac{W'(x')}{x-x'}\omega_{1}(x')
=\omega_{1}(x)\omega_{0}(x)
\ende
which has only the solution $\omega_{1}=0$. This implies that $\widehat{F}_{1}$ is a constant which can
be taken to be $0$.

The next step is to relate the free energy of the modified matrix model $\widehat{F}$ to the free
energy of the unmodified, CS model $F$. Since the leading order loop equation is unchanged,
$\widehat{F}_{0}=F_{0}$. Now we will need to introduce the density function 
$\rho(x)$, it is related to the resolvent by 
\be
\omega(x)=\int_{-a}^{a} \frac{\rho(\lambda)}{x-\lambda}d\lambda. 
\ende
By the saddle point approximation, we can see that

\bea
-\widehat{F}_{1}&=& \frac{1}{2}\int_{-a}^{a} \rho(\lambda){\rm ln}\lambda d\lambda - F_{1} \\
&=& -\frac{1}{2}\int_{-a}^{a} \rho(\lambda)\left( {\rm P}\int^{\Lambda}_{0} \frac{1}{x-\lambda} dx \right) d\lambda
- F_{1}, \\
&=& - \frac{1}{2}\int_{a}^{\Lambda}\omega_{0}(x) dx 
-F_{1}, \\
&=& -\frac{1}{4} \frac{\del F_{0}}{\del S} -F_{1}. \label{F1HMM}
\eea
where $\Lambda$ is some large cutoff. Casting $F_{1}$ as an integral of a 1-form over a Riemann surface
is natural in the context of matrix models with flavour which also has an ${\cal O}(g_{s})$ 
correction to the free energy \cite{Naculich:2002hr}. So (\ref{F1HMM}) implies that 

\be
F_{1}=-\frac{1}{4}\frac{\del F_{0}}{\del S}.
\ende
agreeing with the previous derivation of the same result. Generalizing this 
procedure to the CS matrix model will be the focus of this paper.


\section{Topological String Geometry}

Chern-Simons theory on $S^3$ with the gauge group $SU(M)$ describes 
topological A-branes wrapped around the $S^3$ in the deformed conifold $T^{*}S^3$. 
After the appropriate involution of the $S^3$ conifold geometry \cite{Sinha:2000ap}
the gauge group of the CS theory gets replaced  by $SO(2M)$ or $Sp(2M)$ depending on the sign of the crosscap.
The involution goes through the usual web of dualities and
large N transitions. After the large N transition the closed string geometry is an orientifold of \mbox{$O(-1)+O(-1)\rightarrow \IP^1$},
the $\IP^1$ becoming an $\rp^2$. It is important that the involution does not have any fixed points. The mirror B-model
geometry is again an orientifold of some deformed CY. The involution now has fixed points, two orientifold planes. This geometry has
been also considered in \cite{Acharya:2002ag}. 
It can be viewed as the end point of the large N transition on the B-model side when the B-model branes that are mirror
to the A-model branes on $S^3$ disappear leaving only two orientifold planes. 
This two orientifold planes give a subleading 
contribution $F_1$ to the free energy that is not present for $SU(M)$ gauge group. $F_1$ counts the holomorphic maps of $\rp^2$ into
the resolved conifold.
  
After the canonical quantization the CS theory can be reduced to certain matrix model integrals \cite{Marino:2002fk}.    
Unlike usual Hermitian matrix models where the integration is performed over the Lie algebra measure, 
the integration in the CS matrix model is over the Lie group. The matrix model also can be viewed as a result of canonical quantization of  the 
HCS on the B-model side with a potential that contains double trace terms \cite{Aganagic:2002wv}. 
The mirror of $T^*S^3$ is given by the blownup of

\be\label{mir}
xz=(e^u-1)(e^v-1).
\ende
Clearly, the imaginary $u$ direction is compact, with period $2\pi i$. The appearance of the group measure in the matrix integral 
can be interpreted as a result of the counting of all images of the D-branes in a matrix model with the Lie algebra measure 
\cite{Aganagic:2002wv}

\be
\prod_{n}\prod_{i<j}(u_i-u_j+2\pi in)^2
\sim\prod_{i<j}\sinh^2(\frac{u_i-u_j}{2}).
\ende
If the gauge group is $SO(2M)$, each eigenvalue $u_i$ of the matrix has its partner $-u_i$, so an additional product with the plus
sign between eigenvalues appears in the measure.
 
It is possible to take the planar limit for
this matrix model and obtain a spectral curve, in fact the spectral curve is the nontrivial part of the 
B-model geometry which is obtained after the large N transition.
The leading contribution $F_0$ to the free energy is obtained from the integral of the resolvent over a noncompact (B)
cycle. It is almost trivial to generalize this to $SO(2M)$ matrix model. The main concern of 
this paper will be the subleading part  of the free energy ($F_1$),that is due the 
orientifold planes and is absent for $SU(M)$ case. 
From the closed string point of view, 
the answer must be an integral of a meromorphic one-form from the orientifold planes to some fixed point at infinity. From matrix
model point of view this one-form must be related to the resolvent.

 

\section{Free Energy.}
In this section we review known results for the free energy and relate the parameters of topological string and the matrix 
model. 
The A-model orientifold of the conifold $T^*S^3$ has been considered before. In \cite{Sinha:2000ap} the
partition function of $SO/Sp$ Chern-Simons on $S^{3}$ was calculated to all orders in $g_{s}$ and given a closed
string interpretation. There it was found that $F_{0}^{SO/Sp}=1/2 F_{0}^{SU}$,
i.e.
\be
\label{ff0}
F_{0}^{SO/Sp}(t)=\frac{1}{2}\sum_{n=1}^{\infty}\frac{\e^{-nt}}{n^{3}}.
\ende
The coefficient of $\frac{1}{2}$ is consistent with the orientifold action. From a closed string perspective, $F_{1}$ includes
only the holomorphic maps of $\BR\BP^2$ into the resolved conifold which are odd wrappings, i.e. only $\ZZ_{2}$ equivariant maps contribute
to the instanton expansion. Furthermore the area of an $\rp^2$ instanton is half that of a $S^{2}$ instanton. So
\be
\label{ff1}
F_{1}^{SO/Sp}(t)=\pm \sum_{n \ {\rm odd}}\frac{\e^{-nt/2}}{n^{2}},
\ende
where the $+(-)$ sign is for $SO(Sp)$ respectively. 
In the $SO(2M)$ and $Sp(2M)$ matrix model, we will use a t 'Hooft parameter $S=g_sM$, related to the Kahler modulus $t$, the size of 
blown-up $\rp^2$, by
\be
t=2 S\pm g_s.
\ende
This implies that the following relationship between Chern-Simon's free energy $F^{CS}$ and the
matrix model free energy $F^{MM}$,
\bea
&& F^{MM}_{0}(S)=F^{CS}_{0}(2S) \\
&& F^{MM}_{1}(S)=F^{CS}_{1}(2S)\mp\frac{1}{2}\frac{\del F^{CS}_{0}(2S)}{\del S}.
\eea
So more explicitly we have,

\be
\label{f1matrix}
F_1^{MM}(S)=\pm\left (\frac{1}{2}\sum_{n=1}^\infty\frac{e^{-2nS}}{n^2}+\sum_{n\ {\rm odd}}\frac{e^{-nS}}{n^2}\right )
\ende
with $+(-)$ sign for $SO(Sp)$ respectively. So we immediately see that the derivative relation (\ref{free1}) does not hold for the A-model, this was the observation which motivated the present work.
Note that the matrix model subleading free energy has two pieces. One comes from the nontrivial relation between t 'Hooft parameter S and the Kahler
modulus $t$, the other is the contribution of the orientifold planes.

Now another scenario where ${\cal O}(g_{s}^{1})$ corrections appear is gauge theories with fundamental matter. In \cite{Naculich:2002hr},
the authors found that $F_{1}$ is an integral over the spectral curve of the {\it leading order} 
resolvent $\omega_{0}$, consistent with what one would expect from \cite{Aganagic:2000gs, Aganagic:2001nx}. The fundamental matter shows 
itself in the matrix model as a subleading term in the potential. We will see that the contribution from the orientifold planes also
comes in as a subleading term in the matrix model potential and therefore also can be expressed as an integral of the resolvent.


\section{Loop Equation}
As in \cite{Ashok:2002bi}, we derive a loop equation needed to find the leading and subleading order resolvents,
$\omega_{0}(z)$ and $\omega_{1}(z)$. Since the matrix integral is over the Lie algebra group rather than over 
the Lie algebra the measure factor is different, and as a consequence the expression for the resolvent in terms of
eigenvalues is different. This does not change much in the derivative relation between $\omega_0(z)$ 
and $\omega_1(z)$. 
This relationship
does not however, appear to lift to a nice relationship between $F_{0}$ and $F_{1}$ as (\ref{free1}).
\be
{\cal Z}\sim \int \prod_{i=1}^{M}du_{i}\prod_{j\neq i}{\rm sinh}^{2}(\frac{u_{i}-u_{j}}{2}){\rm sinh}^{2}(\frac{u_{i}+u_{j}}{2})\e 
\left(-\frac{2}{g_s}{\rm W}(u_{i})\right).
\ende
As in the matrix model with measure on the Lie algebra, the integral of the resolvent must be compatible with the 
log of the measure so we define,
\be
\omega(x)\equiv g_{s}\left\langle {\rm Tr} \coth\left(\frac{x-\Phi}{2}\right)  \right\rangle.
\ende
When the group is $SO(2M)$, this becomes,

\be\label{res1}
\omega(x)=g_{s}\left\langle \sum_{i=1}^{M}\left( {\rm coth}(\frac{x-u_{i}}{2}) + {\rm coth}(\frac{x+u_{i}}{2}) \right) \right\rangle .
\ende
It behaves as

\be
\omega(x)=\pm 2S+{\cal O}(x^{-1})\hspace{2cm} x\rightarrow\pm\infty.
\ende
We restrict ourselves to one cut solutions so, we assume that $\omega (z)$ has one cut that runs from $-a$ 
to $a$ along the real axis.
We have defined $S=g_{s}M$. With a potential given by ${\rm W}(x)=\sum_{j=1}^{\infty}\frac{g_{j}}{j}x^{j}$, 
the relationship between the resolvent and the free energy $F$ is

\be \label{res2}
\omega(x)=2S\coth(\frac{x}{2})+\frac{d}{dV(x)}F,
\ende
where the differential operator $\frac{d}{dV(x)}$ can be worked out by Taylor expanding $\coth \left(\frac{x-\Phi}{2}\right)$ around $x$.
The resolvent and the free energy have expansions in $g_{s}$ given by,

\bea\label{expansion}
&&F=\sum_{g,c}g_{s}^{2g+c}F_{2g+c}, \nonumber \\
&&\omega(x)=\sum_{g,c}g_{s}^{2g+c}\omega_{2g+c}(x).
\eea
Combining (\ref{res2}) and (\ref{expansion}), we see that,

\bea\label{exp2}
&& \omega_{0}=\frac{d}{dV(x)}F_{0} + 2S\coth(\frac{x}{2}), \nonumber \\
&& \omega_{j}=\frac{d}{dV(x)}F_{j}, \ \ j>0.
\eea
The loop equation for this model can be derived by demanding reparametrisation invariance of the partition function (details
are in the appendix). It is given by,

\be \label{loop_eqn}
\frac{1}{2}\omega^{2}(x)-g_{s}\coth(x)\omega(x)+2g_{s}S -2S^{2} -{\widehat {\cal K}}\omega(x)+\frac{g_{s}^{2}}{2}\frac{d}{dV(x)}\omega(x)=0. 
\ende
Where ${\widehat {\cal K}}$ acts as,

\be
{\widehat {\cal K}}f(x)=\oint_{\cal C}\frac{dz}{2\pi i}\coth(\frac{x-z}{2}){\rm W}'(z)f(z).
\ende
The contour ${\cal C}$ encircles the cut but not the point $x$.
When we insert the expansion (\ref{expansion}) into the loop equation (\ref{loop_eqn}), the first two equations we get are,

\bea \label{loop2}
{\cal O}(g_{s}^{0}):&& \frac{1}{2}\omega_{0}^{2}(x)-2S^{2}={\widehat {\cal K}}\omega_{0}(x), \\
\label{loop3}
{\cal O}(g_{s}^{1}):&&\omega_{1}(x)\omega_{0}(x)+2S-\coth(x)\omega_{0}(x)={\widehat {\cal K}}\omega_{1}(x).
\eea
From these equations we see that

\be \label{deriv}
\omega_{1}=-\frac{1}{2}\frac{\del \omega_{0}}{\del S}+\coth(x)
\ende
is a solution to (\ref{loop3}) and has the correct behavior at infinity. 
Similar equations for the Lie algebra case were found in \cite{Ashok:2002bi, Janik:2002nz, Alday:2003gb, Kraus:2003jv,
Klemm:2003cy}.
This implies the following relation for the free energy,

\bea \label{deriv_F}
\frac{d}{dV(x)}F_{1}&=&-\frac{1}{2}\frac{\del \omega_{0}}{\del S}+\coth(x) \\
&=& -\frac{1}{2}\frac{d}{dV(x)}\frac{\del F_{0}}{\del S} - \coth(\frac{x}{2}) + \coth(x). \nonumber
\eea
Here the method of \cite{Ashok:2002bi} breaks down. In that situation one could trivially integrate to get $F_{1}$ but
here we are unable to. Doing so would amount to writing $\coth(x)- \coth(\frac{x}{2})$ as $\frac{d}{dV(x)}$
of some function, something we were unable to do. We will derive $F_{1}$ using the same method as we did in the 
section 2.2 for the Lie algebra case. One needs to find a subleading term in the potential, the term that one has
to subtract in order to kill the subleading part $F_1$ of the free energy. The first step to this goal is to find the 
leading order resolvent $\omega_0$

\section{Free Energy of Leading Order.}

To find the resolvent $\omega_0 (z)$ in the loop equation it is easier to go back and derive an equation of motion.
Since we are mostly interested in the Chern-Simons matrix models we again restrict our attention to one cut solutions. 
Let introduce a density function $\rho_{0}(u)$ by,

\be
\omega_0 (z)=g_s\int^a_0\rho_0 (u)\left ( \coth\frac{z-u}{2}+\coth\frac{z+u}{2}\right )du.
\ende
It is convenient to continue the density function to the negative part of real axis \mbox {$\rho (z)=\rho (-z)$}. The above definition becomes

\be
\omega_0(z)=g_s\int^a_{-a}\rho_0 (u)\coth\frac{z-u}{2}du.
\ende
The normalization condition that guarantees the correct behavior of the resolvent at infinity is given by

\be
\label{norm}
\int^a_{-a}\rho_0(u)du=2 S.
\ende
Lets plug this definition into the loop equation for $\omega_0 (z)$ (\ref{loop2}). Subtracting the loop equation evaluated
at the point $z+i\epsilon$ above the cut from the loop equation at the point $z-i\epsilon$ bellow the cut and 
taking into account that

\be
\coth\frac{z-x-i\epsilon}{2}-\coth\frac{z-x+i\epsilon}{2}=4\pi i\delta (z-x),
\ende
one gets

\be
\frac{2}{g_s} W'(z)={\rm P}\int^a_{-a}\rho_0(u)\coth\frac{z-u}{2}du.
\ende
The usual way to proceed is to go to a new coordinate $U'=e^u$ \cite{Aganagic:2002wv}. In the case of $SO(2M)$ 
Chern-Simons matrix model, the potential is $W(z)=z^2/4$ and the equation of motion becomes

\be
\label{eqm}
-\frac{1}{2g_s}\log (Ue^{-2S})={\rm P}\int^{e^a}_{e^{-a}}\frac{\rho_0 (U')}{U'-U}dU',
\ende
where $U=e^z$. Here the normalization condition 

\be
g_s\int^{e^a}_{e^{-a}}\rho_0 (U')\frac{dU'}{U'}=2 S
\ende
has been used. Following \cite{Aganagic:2002wv, Kazakov:1995ae} it is easy to find the function

\be
v (U)=g_s\int^{e^a}_{e^{-a}}\frac{\rho_0 (U')}{U'-U}dU',
\ende
that satisfies the following

\bea
&& i)\  {\rm vanishes\ at\ infinity}, \nonumber \\
&& ii)\ {\rm has\ a\ square\ root\ cut}, \nonumber\\
&& iii)\ v(0-i\epsilon)=2 S, \nonumber\\
&& iv)\ v(U-i\epsilon)+v(U+i\epsilon)=-1/g_s\log (Ue^{-2S}).\nonumber 
\eea
The only difference from the $SU(M)$ case considered in \cite{Aganagic:2002wv} is
that $S$ gets doubled,

\be
v(U)=\log\frac{1+U+\sqrt{{(1+U)}^2-4Ue^{2S}}}{2U}.
\ende
The relationship between this function and the resolvent is 

\be
\omega_0(z)=2S-2v(e^z).
\ende

Let's discuss the geometry of the Riemann surface given by the resolvent $v (u)$. The spectral curve that corresponds to the resolvent is
given by
\be
\label{Rsurface}
e^{v}-e^{-u}+e^{-u-v+2S}-1=0.
\ende
Since the resolvent has the property $v(u)=v(u+2 \pi i)$ the Riemann surface is compact in the imaginary direction. The resolvent has
the square root cut giving rise to the two sheets of the surface. Therefore the Riemann surface looks like two infinite cylinders
glued together along the cut.  The contour around the
cut is usually called an A cycle, the contour running from a point at infinity on one sheet to a point at infinity on the other
sheet is called a B cycle. The Riemann surface is depicted in fig. \ref{graph} and fig. \ref{shtuka}.

From the string theory point of view the curve is part of the mirror B-model geometry\cite{Acharya:2002ag}. The branes have 
disappeared. In terms of type IIA strings
there are two orientifold O5-planes at points $u=\pm i\pi$. These will play a role in the next section when we calculate the
${\cal O}(g_{s})$ part of the free energy. From the string theory point of view,
the O5-plane contribution to the superpotential is an integral from the location of the O5 plane to the point at infinity. The two O5-planes have different D-brane
charges therefore their contributions are summed with opposite signs.  We will see how these results emerge from the matrix model.   

Now we are ready to calculate the leading contribution $F_0(S)$ to the free energy. As usual it is given by the integral of 
the resolvent over the $B$ cycle. The integral can be expressed in terms of Euler's dilogarithm function (see appendix B for
definition and useful properties)

\be
\label{f0}
\partial_S F_0(S)=\frac{1}{2}\int_B\omega_0 (z)dz=-\int_Bv(U)\frac{dU}{U}=-\Li\left (e^{-2S}\right ).
\ende
Here all infinite and polynomial in $S$ terms are omitted, leaving the worldsheet instanton contribution,

\be
F_0(S)=\frac{1}{2}\sum^\infty_{n=1}\frac{e^{-2nS}}{n^3},
\ende
which agrees with \cite{Sinha:2000ap}.
The next objective is to calculate $F_1$ and cast it into the form of

\section{Free Energy of Order ${\cal O}(g_s)$}
The next objective is to calculate ${\cal O}(g_s)$ contribution to the free energy. Although there is a derivative relation between
$\omega_0(z)$ and $\omega_1(z)$ (\ref{deriv}), this cannot be integrated to a relation between $F_{0}$ and $F_{1}$ (\ref{deriv_F}).
Therefore to find the ${\cal O}(g_{s})$ part of the free energy we use the new method that we have described in the second part of section 2.

\subsection{$F_1$ as a dilogarithm}

To do so we consider the origin of the ${\cal O}(g_{s})$ term in the free energy. Since the saddle point method is 
used to construct the perturbative expansion in powers of $g_s$, there should not be any terms of order of $g_s$ unless
there is a subleading term in the effective matrix model action. 
To single out such a piece, we add the following subleading term to the action,

\be\label{delW}
{\rm Tr}\delta W (u)=\frac{1}{2}\sum_i\log\sinh^2 u_i.
\ende
This is analogous to (\ref{addW}) in the case of the Lie algebra matrix model.
For this new potential, $W+\delta W$, we denote the free energy $\widehat{F}_1$.
The equation for $\omega_0 (z)$ (\ref{loop2}) is invariant but the equation for $\omega_1(z)$ (\ref{loop3}) becomes
\be
{\widehat {\cal K}}(z)\widehat{\omega}_1=\omega_0(z)\widehat{\omega}_1(z).
\ende
Provided that $\widehat{\omega}_1(z)$ vanishes as $z\rightarrow\pm\infty$ this integral equation has only the trivial solution
$\widehat{\omega}_1=0$, which leads to $\widehat{F}_1=0$. This suggests that (\ref{delW}) cancels the subleading
part of the action. So, similar to (\ref{F1HMM}), we have,

\be \label{addfree}
0=-\widehat{F}_1=-F_1+\frac{1}{2}\int^a_0\rho_0(z)\log\sinh^2 z dz.
\ende
In principle, if one knows the density function $\rho_0(z)$, the integral can be taken. It is more convenient however
to have it written as an integral of $\omega_0(z)$, which is known.
We will see that written in that form, $F_1$ has two different pieces, corresponding to the integration along
different contours. 

It is easy to see that

\be
\log\sinh z=\log 2+\log\sinh\frac{z}{2}+\log\sinh\frac{z+i\pi}{2}-\frac{i\pi}{2}.
\ende

\begin{figure}[bt]
\centerline{ \epsfig{file=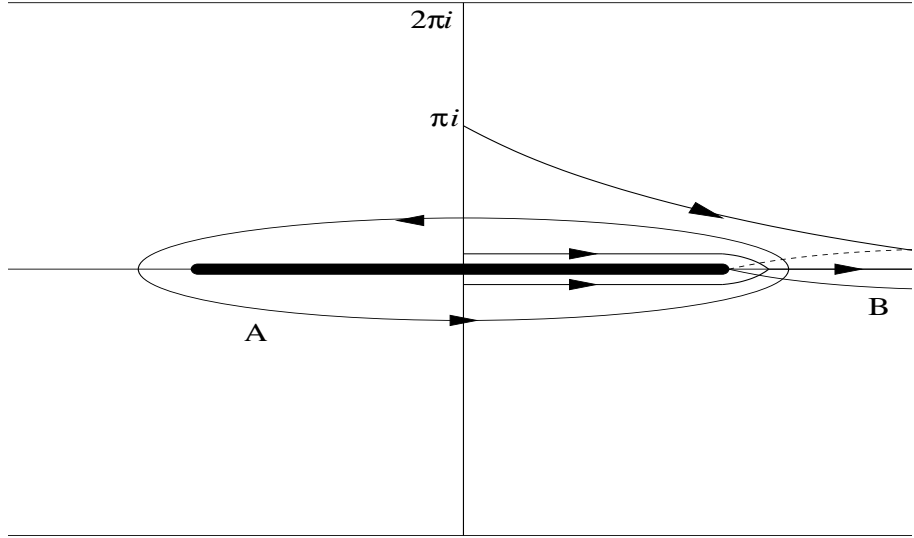,width=12cm,height=12cm}}
\caption{\sl The physical sheet of the spectral curve.}
\label{graph}
\end{figure}

In what follows we omit all infinite terms and polynomial terms that can be easily restored.
Similar to \cite{Naculich:2002hr}, we write the logarithms in the integral (\ref{addfree}) as 

\be
\log\sinh\frac{z}{2}=-\frac{1}{2}{\rm P}\int^\Lambda_0\coth\frac{x-z}{2}dx-\frac{z}{2}
\ende

\be
\log\sinh\frac{z+i\pi}{2}=-\frac{1}{2}\int^\Lambda_{i\pi}\coth\frac{x+z}{2}dx+\frac{z}{2}
\ende
where $\Lambda$ is a point at infinity, a UV cutoff. Now one can 
recognize the resolvent in (\ref{addfree}). Combining the above and using the fact that
$\rho_0 (z)$ is an even function we get

\be
\label{F1} 
F_1 (S)=-\frac{1}{4}\left \{{\rm P} \int^\Lambda_0\omega_0 (z)dz+\int^\Lambda_{i\pi}\omega_0 (z)dz\right \}.
\ende
To figure out which branch of the function $\omega_0(z)$ is to be used in the above integrals or in other
words on which sheet the point $\Lambda$ is located, one has to look at the behavior at infinity of 
the integrals of the $\coth(z/2)$ function. The conclusion is that $\Lambda$ is a point on the physical
sheet where $\omega_0(z)$ vanishes at infinity.
The principal value integral is to be understood as ${\rm P}\int_0=1/2 (\int_{0+i\epsilon}+\int_{0-i\epsilon})$.
The two contours are drawn on fig. \ref{graph}.
So far our discussion in this section has been applicable to one cut solutions of the matrix model with arbitrary 
potentials. From this point we restrict ourselves to quadratic potentials only.

The first integral in (\ref{F1}) can be written as an integral
over the B cycle. To see this, we introduce a function $y(z)$ that corresponds to the 
singular part of the resolvent, so the integrals over a cycle of $y(z)$ or of the resolvent are the same. 
The function $y(z)$ has the property of a square root, i.e. it changes the sign when passing
over the cut, $y(z+i\epsilon)=-y(z-i\epsilon)$. The differential $y(z)dz$ is the meromorphic one-form
on the Riemann surface defined by the resolvent. One has  

\be
\omega_0(z)=-2 W'(z)+y(z),
\ende

\be
-\frac{1}{4}{\rm P}\int^\Lambda_0\omega_0(z)dz=-\frac{1}{8}\int_B\omega_0(z)dz+\frac{1}{2}W(\Lambda).
\ende
To check the result (\ref{F1}) we calculate $F_1(S)$ for the quadratic potential and compare it with the
known result. The integral over the B cycle has been taken already in the previous section. Therefore

\be
\label{f11}
-\frac{1}{4}{\rm P}\int^\Lambda_0\omega_0(z)dz=\frac{1}{4}\Li\left (e^{-2S}\right ).
\ende   
The second integral in (\ref{F1}) can also be expressed in terms of Euler's dilogarithm function 
(see Appendix B for the details),

\be
\label{final}
-\frac{1}{4}\int^\Lambda_{i\pi}\omega_0(z)dz=\frac{1}{4}\Li (e^{-2S})+\frac{1}{2}\left (
\Li (e^{-S})-\Li(-e^{-S})\right ).
\ende
where again we dropped all polynomial terms. 
After summing (\ref{f11}) and (\ref{final}) one gets the correct free energy $F_1(S)$ (\ref{f1matrix}).
\begin{figure}[bt]
\centerline{ \epsfig{file=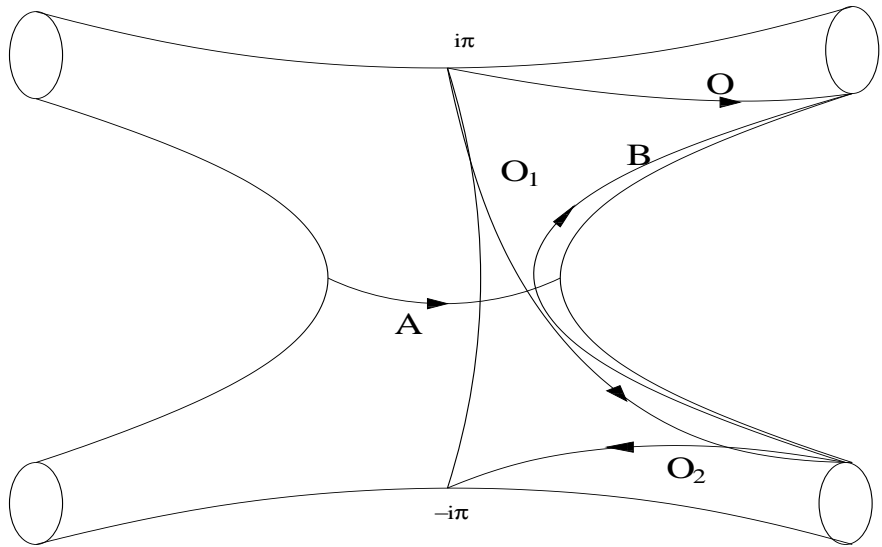,width=12cm,height=12cm}}
\caption{\sl The Riemann surface and and the integration contours that give ${\cal O}(g_{s})$ part of the free energy.}
\label{shtuka}
\end{figure}


\subsection{$F_{1}$ as a contour integral}
There is another way of writing (\ref{final}) which makes contact with the results of \cite{Acharya:2002ag}.
We want to separate clearly the part coming from O5-planes
and the part from the expansion of $F_0(t)$. The last one is just the integral over the B cycle. Again it is convenient to 
appeal to the meromorphic one-form $y(z)dz$

\be
\int^\Lambda_{i\pi}\omega_0(z)dz=\frac{1}{2}\left (\int^\Lambda_{i\pi}y(z)dz+\int^{-i\pi}_{\Lambda '}y(z)dz\right ).
\ende
The contour in the second integral is on the second sheet of the Riemann surface, and 
$\Lambda '$ is a point at infinity on this sheet (see fig. \ref{shtuka}). Let denote the contour in 
the first integral as $O$ and in the second as $O_2$. From figure \ref{shtuka} it is 
clear that $O=O_1+B$ where the contour $O_1$ goes from $i\pi$ to $\Lambda '$. The conclusion is 

\be
\int^\Lambda_{i\pi}\omega_0(z)dz=\frac{1}{2}\left (\int_B\omega_0(z)dz+\int_{O_1}\omega_0(z)dz+\int_{O_2}\omega_0(z)dz\right )
\ende
The contour $O_1$ and the contour $O_2$ run from the positions of two O5-planes to the point at infinity which is precisely what
is expected from the string theory point of view. This consideration so far is valid for arbitrary potentials. If the potential
is quadratic one has to take integrals of $v(u)$. The change of variables $v(u)du=vu'(v)dv$ brings the integrals over 
$O_1$ and $O_2$ contours to the same form as in \cite{Acharya:2002ag}.

Almost the same calculation can be done for the case of $Sp(2M)$ group. The group measure for $Sp(2M)$ has the extra
factor $\prod_i\sinh^2 u_i$ which corresponds to the additional term $\sum_i\log\sinh^2u_i$ in the matrix model
effective action. To have $\tilde{F}_1=0$ the term 

\be
{\rm Tr}\delta {\rm W} (u)=-\frac{1}{2}\sum_i\log\sinh^2u_i
\ende 
has to be added to the potential. It is clear now that the only difference from $SO(2M)$ case is the opposite sign of $F_1(S)$ 
which agrees with (\ref{f1matrix}).

While it is possible
to continue the Chern-Simons partition function to $SO(2M+1)$
\cite{Sinha:2000ap} and take the large $M$ limit,
this partition function does not match the partition function of the 
$SO(2M+1)$ matrix model.
So although the $SO(2M+1)$ matrix model does not corresponds to Chern-Simons
theory one can still consider this model. 
The group measure has an extra factor $\prod_i\sinh^2\frac{u_i}{2}$. Repeating
the above procedure one gets

\be
F_1=\frac{1}{2}\sum_i\left (\log 2-\frac{i\pi}{2}+\log\sinh^2\frac{u_i+i\pi}{2}
-\log\sinh^2\frac{u_i}{2}\right ).
\ende
Since the second logarithmic function has changed sign compared to the 
$SO(2M)$ case the two integrals over the B cycle cancel each other and we are 
left with integrals over the contours $O_1$ and $O_2$. So the part that is 
proportional to the derivative of $F_0$ disappears. This is in contrast with 
the Lie algebra case, in which the $SO(2M+1)$ and $Sp(2M)$ matrix
models have the same free energy.

\section{{\cal N=1} SYM In Four Dimensions}

The tree level superpotential is figured out (following \cite{Marino:2002fk}) by converting the Haar measure into a measure
on the Lie Algebra

\bea
&&\prod_{i < j}\left(  2\sinh \frac{u_{i}-u_{j}}{2}\right)^{2} 
\left(  2\sinh \frac{u_{i}+u_{j}}{2}\right)^{2} \\
&&=\prod_{i < j}(u_{i}^{2}-u_{j}^{2})^{2} {\rm exp}\left( \sum_{k=1}^{\infty} 
a_{k}\left( \sigma_{k}^{-}+\sigma_{k}^{+}\right) (u) \right),
\eea
where,

\be
\sigma_{k}^{\pm}(u)=\sum_{i<j}(u_i \pm u_j)^{2k}, \ \ \ \ \ \ a_{k}=\frac{B_{2k}}{k(2k)!}, \nonumber
\ende
$B_{2k}$ are the Bernoulli numbers. Using the fact that the Newton polynomials

\be
P_{k}(u)=\sum_{i=1}^{M}u_{i}^{k},
\ende
are equal to $\frac{1}{2}{\rm Tr}M^{k}$, where $M$ is an anti-symmetric matrix gauge fixed to the diagonal,
we see that the four dimensional tree-level superpotential engineered by this construction
is 

\bea \label{wtree}
&&W_{\rm tree}(\Phi)=  \\
&&\frac{1}{8}\Phi^{2}-\frac{g_{s}}{2}\sum_{k=0}^{\infty}a_{k}
\left[-2^{2k-2}\Phi^{2k}+
\sum_{s=0}^{2k}
\left(\begin{array}{c} 2k \\ s \end{array}\right)
\frac{1}{4N}{\rm Tr}\Phi^{s}{\rm Tr}\Phi^{2k-s}(1+(-1)^{s})
\right]. \nonumber
\eea
Only even powers of $\Phi$ appear, as expected.

The Chern-Simon's partition function will give the four dimensional low energy Wilsonian
effective action for ${\cal N}=1$ SYM with $W_{\rm tree}$ given by (\ref{wtree}). Now in \cite{Sinha:2000ap}, the correct 
closed string variable was identified as $t=g_{s}(2M-1)$ (for $SO(2M)$), which is identified
with the gluino condensate. So following \cite{Sinha:2000ap, Ashok:2002bi} we propose the formula

\be\label{superpot}
W_{\rm eff}={\cal Q}_{D6}\frac{\del F_{0}^{CS}}{\del t}+{\cal Q}_{O6}G_{0}^{CS}-\tau t,
\ende
where ${\cal Q}_{D6}$ is the total D6-brane charge and ${\cal Q}_{O6}$ is the total O6-plane charge, $\tau$ is the gauge coupling.
$Op$-plane charge is given by $\pm2^{5-p}$, -sign for $SO$, +sign for $Sp$. We have introduced $G_{0}=aF_{1}^{CS}$, $a$ is a constant.
It is important to use $F^{CS}$ not $F^{MM}$ in \ref{superpot} because their $g_{s}$ expansions differ, as discussed in the introduction.

To make contact with results of \cite{Witten:1997bs}, we look at ${\rm log(vol(}SO(2M))$. This term is already within 
the free energy but we know from \cite{Ooguri:2002gx} that it is this term which supplies the $t\log t$ term to the
superpotential. So, expanding in $(2M-1)$ and keeping only log terms, we get 

\be
-\log ({\rm vol}(SO(2M))\sim g_{s}^{-2}\frac{t^{2}}{4}\log t + g_{s}^{-1}\frac{t}{4}\log t + \ldots
\ende 
so using (\ref{superpot}), we find 

\be
W_{\rm eff}=\frac{N}{4}t\log t-\frac{1}{2}\frac{at}{4}\log t + \frac{N}{2}\frac{\del F^{pert}_{0}}{\del t}-2F_{1}^{pert}-\tau t,
\ende
where $F^{pert}=F+\log ({\rm vol}(SO(2M))$. So requiring $N-2$ vacua, we find that $a=4$.
\section{Conclusion}
We have studied the matrix models with Haar measure on $SO/Sp$ in the large M limit, for which we have introduced the new form 
of the resolvent that is compatible with the group measure.
We have derived a quantum loop equation and for the case of quadratic potential, and
have found the leading order resolvent.

We have calculated the
${\cal O}(g_s)$ corrections to the $SO$ and $Sp$ Gaussian matrix model free energy using a novel method. This method separates in 
a clear way the leading (of order $1/g_s$) and subleading (of order ${\cal O}(1)$) parts in the effective 
matrix model action. The free energy of the first two orders was expressed as integrals of the leading order
resolvent over the spectral curve and these integrals were explicitly performed.
 
We have found agreement between matrix model and large M Chern-Simons results. While the $g_{s}$ expansion 
of the Chern-Simons theory has a nice worldsheet interpretation, which means that the first two orders correspond to sphere and 
$\rp^{2}$ worldsheets, the $g_{s}$ expansion of the matrix model free energy mixes the worldsheet contribution
at each order, essentially due to a shift in the identification of the 't Hooft parameter. The derivative relation found in
Lie algebra matrix models

\be
F_{1}=-\frac{1}{4}\frac{\del F_{0}(S)}{\del S}
\ende
does not hold but instead we find that

\be
F_{1}=-\frac{1}{2}\left (\frac{\del F_{0}(S/2+i\pi)}{\del S}+\frac{\del F_{0}(S/2)}{\del S}+
\frac{\del F_{0}(S)}{\del S}\right ).
\ende
Here $F_{1}$ contains a contribution from sphere worldsheets as well as $\rp^{2}$ worldsheets.

Type IIA string theory on $M^{3,1}\times T^{*}S^{3}$ with the internal geometry orientifolded and $N$ D6 branes wrapped on $M^{3,1}\times S^{3}$ engineers
an ${\cal N}=1$ $SO/Sp\ $ SYM in four dimensions with a certain double trace tree level superpotential which was given.
The calculation of the leading and subleading free energy in these matrix models or equivalently in Chern-Simons
theory gives the effective superpotential for this four dimensional SYM. This was also discussed.

Although we have presented the main results with the potential $W(z)$
an arbitrary polynomial, the string theory application of these 
matrix models is known only for the quadratic potential. One can convert
the matrix model to a Lie algebra matrix model with  double trace 
potential and a single trace potential $W(\Phi)$. This potential then corresponds to 
the tree level superpotential of an ${\cal N}=1$ SYM in four dimensions. Therefore one knows 
the four dimensional effective theory but does know the internal geometry which
constructs this theory. 
If the potential has higher than quadratic powers, the spectral curve will not be a polynomial in $e^u$ and $e^v$.
Whether or not this spectral curve can be related to some B-model geometry is an
interesting question to address. 

It would be interesting to generalize the impressive work \cite{Tierz:2002jj} and solve the $SO/Sp$ matrix model
to all orders by the method of orthogonal polynomials.


\begin{acknowledgments}
We are particularly grateful to Mina Aganagic for bringing \cite{Naculich:2002hr} to our attention. We also thank
Marcos Marino for explaining his work to us.
Thanks are due to Nick Warner for the reading the manuscript.
We have much enjoyed the interaction with all members of the string group
at USC. 
NH is supported by a Fletcher Jones Graduate Fellowship.
\end{acknowledgments}

\begin{appendix}

\section{Derivation Of the Loop Equation}

In this paper we are just interested in calculating the free energy contribution from
$\rp^2$ worldsheets. To do this we derive an all worldsheet loop equation. This equation will 
also be derived for Lie group $SU(M)$ for completeness.

\subsection{$SO(2M)$}
The partition function, after integrating out the off diagonal terms, is given by,

\be
{\cal Z}\sim \int \prod_{i=1}^{M}du_{i}\prod_{j\neq i}{\rm sinh}^{2}(\frac{u_{i}-u_{j}}{2}){\rm sinh}^{2}(\frac{u_{i}+u_{j}}{2})\e 
\left(-\frac{2}{g_s}{\rm W}(u_{i})\right).
\ende
We perform the infinitessimal change of co-ordintes, 

\be
u_{i}\rightarrow u_{i} + \epsilon\left( {\rm coth}(\frac{x-u_{i}}{2}) - {\rm coth}(\frac{x+u_{i}}{2})\right).
\ende
%
and demand that the partition function is invariant under this transformation. At first order this yields the following constraint,

\bea\label{loop1}
&& \sum_{i=1}^{M}\left\langle \frac{1}{2}\left( \cosech^{2}(\frac{x-u_{i}}{2}) + \cosech^{2}(\frac{x+u_{i}}{2})\right)\right. \nonumber\\
&& +\sum_{j\neq i}\left(  \coth(\frac{x-u_{i}}{2}) - \coth(\frac{x+u_{i}}{2}) \right)
\left(  \coth(\frac{u_{i}-u_{j}}{2}) + \coth(\frac{u_{i}+u_{j}}{2}) \right)\nonumber \\
&& \left. -\frac{2}{g_s}\left( \coth(\frac{x-u_{i}}{2}) - \coth(\frac{x+u_{i}}{2}) \right){\rm W}'(u_i)\right\rangle=0.
\eea

Now we need the 2 identities,

\bea \label{id1}
&&\sum_{i,j=1}^{M}\frac{\cosh(\frac{u_i-u_j}{2})}{\sinh(\frac{x-u_{i}}{2})\sinh(\frac{x-u_{j}}{2})} \nonumber \\
&& =\sum_{i=1}^{M}\cosech(\frac{x-u_{i}}{2})\cosech(\frac{x-u_{j}}{2})
+2\sum_{i\neq j}\coth(\frac{x-u_{i}}{2})\coth(\frac{u_i-u_{j}}{2})\nonumber \\
&& = \sum_{i,j=1}^{M}\coth(\frac{x-u_{i}}{2})\coth(\frac{x-u_{j}}{2})-M^2
\eea

and 

\bea\label{id2}
&&\sum_{i,j=1}^{M}\frac{\cosh(\frac{u_i-u_j}{2})}{\sinh(\frac{x+u_{i}}{2})\sinh(\frac{x+u_{j}}{2})} \nonumber \\
&&=\sum_{i=1}^{M}\cosech(\frac{x+u_{i}}{2})\cosech(\frac{x+u_{j}}{2})
-2\sum_{i\neq j}\coth(\frac{x+u_{i}}{2})\coth(\frac{u_i-u_{j}}{2})\nonumber \\
&& = \sum_{i,j=1}^{M}\coth(\frac{x+u_{i}}{2})\coth(\frac{x+u_{j}}{2})-M^2.
\eea

With a little more work, one can also show that


\bea\label{id3}
&& \frac{1}{2}\omega^{2}(x)-g_{s}\coth(x)\omega(x)+g_{s}S \nonumber \\
&& =\frac{g_{s}^{2}}{2}\sum_{i,j=1}^{M}\left[
\coth(\frac{x+u_{i}}{2})\coth(\frac{x+u_{j}}{2})+\coth(\frac{x-u_{i}}{2})\coth(\frac{x-u_{j}}{2})\right] \nonumber \\
&& +g_{s}^{2}\sum_{i\neq j}\left(\coth(\frac{x-u_{i}}{2})-\coth(\frac{x+u_{j}}{2})\right)\coth(\frac{u_i + u_j}{2}) + S^{2}-g_{s}M.
\eea

%
%
%
Now we multiply (\ref{loop1}) through by $g_{s}^{2}$ and employ (\ref{id1}, \ref{id2}, \ref{id3}) to get the final form of the loop equation,

\bea \label{loop_final}
&&\frac{1}{2}\omega^{2}(x)-g_{s}\coth(x)\omega(x)+2g_{s}S -2S^{2}-{\widehat {\cal K}}\omega(x)+\frac{g_{s}^{2}}{2}\frac{d}{dV(x)}\omega(x)=0, 
\eea
where the linear operator $\widehat{\cal K}$ acts as 

\be
{\widehat {\cal K}}f(x)=\oint_{\cal C}\frac{dz}{2\pi i}\coth(\frac{x-z}{2}){\rm W}'(z)f(z).
\ende

\subsection{$SU(M)$}
Just as above, we integrate out the off diagonal components and demand reparametrisation invariance under
\be
u_{i}\rightarrow u_{i} + \epsilon{\rm coth}(\frac{x-u_{i}}{2}).
\ende
Since the derivation is much simpler than for $SO(2M)$, we just state the result,

\be
\frac{1}{2}\omega^{2}(x) -2S^{2}-\frac{1}{2}{\widehat {\cal K}}\omega(x)+\frac{g_{s}^{2}}{2}\frac{d}{dV(x)}\omega(x)=0. 
\ende


\section{Dilogarithm Identities}

Euler's dilogarithm function $\Li$ is defined as the integral

\be
\Li (z)=\int^0_z\frac{\log (1-t)}{t}dt,
\ende
or as the power series

\be
\Li (z) =\sum^\infty_{n=1}\frac{z^k}{k^2}.
\ende

Among the many amazing properties of this function, we will use the following (for a review see for example \cite{Kirillov:en})

\bea \label{1}
&&\Li (z) +\Li(-z)=\frac{1}{2}\Li (z^2), \\
\label{2}
&&\Li (1-z)+\Li (1-z^{-1})=\frac{1}{2}\log^2 z, \\
\label{3}
&&\Li (z)+\Li (1-z)=\frac{\pi^2}{6}-(\log z)(\log (1-z)). 
\eea

The integral from $i\pi$ to $\Lambda$ can be taken, with the result

\bea
\frac{1}{4}\int^{\Lambda}_{i\pi}\omega_0 (z)dz &=&-\frac{1}{2}\Big (
\frac{1}{2}\log^2 \left (-e^{-S}\right )-\log \left (-e^S\right )
\log \left (1+e^{-S}\right ) \nonumber \\
&&+\Li \left (e^{-2S}\right )-\Li \left (-e^{-S}\right )
-\Li \left (1+e^S\right )\Big ).
\eea
To cancel the product of logarithms one has to apply (\ref{2}) to $\Li\left (1+e^S\right)$
to get $\Li \left (1+e^{-S}\right )$, then using (\ref{3}) convert it to $\Li \left (-e^{-S}\right )$.
After this manipulation and omitting all polynomial terms in S, one has

\be
\frac{1}{4}\int^\Lambda_{i\pi}\omega_0 (z)dz=-\frac{1}{2}\left ( -2 \Li \left (-e^{-S}\right )
+\Li \left (e^{-2S}\right )\right ).
\ende
The last step is to apply the property (\ref{1}) and get (\ref{final}). 
\end{appendix}

\end{document}